\documentclass[aps,prd,a4paper,twocolumn,amsmath,showpacs,superscriptaddress,nofootinbib,preprintnumbers]{revtex4-1}

\usepackage{graphicx,ulem}
\usepackage{longtable}
\usepackage{float}
\usepackage{dcolumn}
\usepackage{graphics,epsfig}
\usepackage{amsmath,amssymb,latexsym,mathrsfs}
\usepackage{bm}
\usepackage{color}
\usepackage{color}
\usepackage{subfigure}

\def\he4{$^4$He}
\def\h2{$^2$H}

\def\aap{\ref@jnl{A\&A}}                

\begin{document}
\title{Recovering a MOND-like acceleration law in mimetic gravity}

\author{Sunny Vagnozzi}
\email{sunny.vagnozzi@fysik.su.se}
\affiliation{The Oskar Klein Centre for Cosmoparticle Physics, Stockholm University, Roslagstullbacken 21A, SE-106 91 Stockholm, Sweden}

\date{\today}

\begin{abstract}
We reconsider the recently proposed mimetic gravity, focusing in particular on whether the theory is able to reproduce the inferred flat rotation curves of galaxies. We extend the theory by adding a non-minimal coupling between matter and mimetic field. Such coupling leads to the appearance of an extra force which renders the motion of test particles non-geodesic. By studying the weak field limit of the resulting equations of motion, we demonstrate that in the Newtonian limit the acceleration law induced by the non-minimal coupling reduces to a Modified Newtonian Dynamics (MOND)-like one. In this way, it is possible to reproduce the successes of MOND, namely the explanation for the flat galactic rotation curves and the Tully-Fisher relation, within the framework of mimetic gravity, without the need for particle dark matter. The scale-dependence of the recovered acceleration scale opens up the possibility of addressing the missing mass problem not only on galactic but also on cluster scales: we defer a full study of this issue, together with a complete analysis of fits to spiral galaxy rotation curves, to an upcoming companion paper. 
\end{abstract}


\maketitle

\section{Introduction: the galactic rotation curves problem}
\label{rotationcurves}

Over the past years, a plethora of cosmological and astrophysical observations have provided astounding confirmation of the ``dark Universe" picture, which by now represents one of the cornerstones of modern cosmology (see e.g. \cite{2df,act,spt,wigglez,wmap,boss,bicep,planck}). Accurate data ranging from CMB measurements to galaxy redshift surveys and supernovae surveys support a concordance cosmology model wherein ordinary baryonic matter only accounts for $\sim 4\%$ of the energy budget of the Universe. On the other hand $\sim 24\%$ of the energy content of the Universe is presumed to be in the form of non-baryonic dark matter, responsible for the formation of structures we observe today. Finally, the remaining $\sim 76\%$ is dubbed dark energy, and is assumed to be driving the late-time accelerated expansion of the Universe. \\

The existence of dark matter (DM) was first inferred by the pioneering works of Oort \cite{oort} and Zwicky \cite{z,zwicky} in the 1930s. However, it wasn't until the 1970s that the community started taking the idea of non-baryonic dark matter seriously, following the work of Rubin, Ford, and Thonnard \cite{rubin,ford}, who measured rotation curves of edge-on spiral galaxies with unprecedented accuracy. If the luminous matter content of galaxies were the sole responsible for the inferred rotation curves, simple Newtonian mechanics considerations dictate that such curve would fall off with radius $r$ as $v_{\text{rot}} \propto 1/\sqrt{r}$. However, the works of Rubin \textit{et al.} conclusively proved that such rotation curves were in fact to good approximation flat, i.e. $v_{\text{rot}} \approx {\rm const}$, far beyond the edge where luminous matter is present. From this observation, one is led to the conclusion that the matter content of the galaxy is not limited to that of luminous matter, but encompasses an additional dark component (the dark matter) whose mass distribution is such that $M(r) \propto r$, responsible for the inferred flat rotation curves. \\

Several candidates for DM exist in the literature (for reviews see e.g. \cite{bertone,primer,gelmini}), with many models positing the existence of additional particles and forces other than those accounted for by the Standard Model of Particle Physics (e.g. \cite{feng,ibarra,arkanihamed,kaplan,cline,cyrracine,fan,foot,petraki,vagnozzi,vagnozzi1,
roxlo,vagnozzi2}). However, several of these models are constructed \textit{ad hoc} in order to explain unexpected experimental results (notably possible signals of annihilating DM), whereas more ``natural" scenarios (such as the SUSY WIMP) are being put under serious pressure by the persisting lack of detection of DM in current experimental efforts (such as direct detection experiments, indirect detection experiments, and collider searches). \\

It is possible, however, that DM might not consist of a particle, but instead be the manifestation of a theory of gravity  beyond General Relativity. In recent years, modified theories of gravity have gained considerable interest in the community: within such theories, dark matter and dark energy often naturally appear as geometrical effects. An example of such theory is known as Modified Newtonian Mechanics (MOND) \cite{milgrom}, where the behaviour of the gravitational force deviates from that of Newtonian mechanics at low accelerations. Another thoroughly explored theory beyond General Relativity where dark matter can appear as a geometrical effect is that of $F(R)$ gravity, which can also naturally unify inflation and the late-time acceleration (see e.g. \cite{cenosz}, for reviews refer for instance to \cite{odintsovreview1,faraoni,defelice,odintsovreview2,sebastiani}). For more general reviews on modified theories of gravity, or concerning theories not discussed here, see e.g. \cite{odintsovreview3,clifton,bamba,mg,cai,dmhorndeskirinaldi}. \\

It is with a particular modified theory of gravity, namely mimetic gravity \cite{mimetic1,mimetic2} (see also the recent review~\cite{svm}), and the possibility of solving the galactic rotation curves problem therein, that this \textit{Letter} shall be concerned. This \textit{Letter} is intended as in ideas paper, wherein we propose a theoretical basis for explaining flat rotation curves in mimetic gravity. In a companion paper to appear subsequently, we will test the idea by providing a fit to rotation curve data \cite{mine}.~\footnote{We note that there currently exist no observational constraints on mimetic gravity from astrophysical and cosmological data neither in the weak nor in the strong field regime.} \\

The rest of this \textit{Letter}, then, is structured as follows. In Section \ref{mimeticgravity}, we briefly review mimetic gravity. Subsequently, in Section \ref{nm} we extend such theory by adding a non-minimal coupling between the matter sector and the mimetic field. Section \ref{nl} is devoted to the examination of the Newtonian limit of the theory. We show that the acceleration law reduces to that of a MOND-like theory, which leads us to examine implications for explaining rotation curves. Finally, Section \ref{conclusion} contains concluding remarks.

\section{Mimetic gravity}
\label{mimeticgravity}

The term ``mimetic dark matter" (later referred to as mimetic gravity) was coined in 2013 by Chamseddine and Mukhanov \cite{mimetic1} (see also \cite{mimetic2}), although the prototype for mimetic theories had actually been introduced a few years earlier in \cite{vikman,gao,capozziello}. In this approach, the conformal degree of freedom of gravity is isolated covariantly, by parametrizing the physical metric $g_{\mu \nu}$ in terms of an auxiliary metric $\tilde{g}_{\mu \nu}$ and a scalar field (the ``mimetic field") $\phi$, as follows:
\begin{eqnarray}
g_{\mu \nu} = \tilde{g}_{\mu \nu}\tilde{g}^{\alpha \beta}\partial_{\alpha}\phi\partial_{\beta}\phi \, .
\label{mimetic}
\end{eqnarray}
From Eq.(\ref{mimetic}) it immediately follows that, for consistency, the mimetic field is required to satisfy:
\begin{eqnarray}
g^{\mu \nu}\partial_{\mu}\phi\partial_{\nu}\phi = 1 \, .
\label{mimeticc}
\end{eqnarray}
With the parametrization in Eq.(\ref{mimetic}), the physical metric is invariant under conformal transformation of the auxiliary metric of the type: $\tilde{g}_{\mu \nu} \rightarrow \Omega(t,\mathbf{x})^2\tilde{g}_{\mu \nu}$. This suggests that mimetic gravity can be viewed as a conformal (Weyl-symmetric) extension of General Relativity, a fact that was first pointed out in \cite{barvinsky}. \\

The gravitational field equations are obtained by varying the Einstein-Hilbert action with respect to the physical metric, taking into account the dependence of the latter on the auxiliary metric and the mimetic field, which yields \cite{mimetic1}:
\begin{eqnarray}
G_{\mu \nu} = \kappa^2 T_{\mu \nu} + \kappa^2 (G - T)\partial_{\mu}\phi\partial_{\nu}\phi \, .
\label{eo}
\end{eqnarray}
where $\kappa^2 \equiv 8\pi G_N$, $G_N$ being Newton's constant. Compared to the conventional Einstein equations, i.e. $G_{\mu \nu} = \kappa^2 T_{\mu \nu}$, Eq.(\ref{eo}) features an additional contribution on the right-hand side, which can be interpreted as the energy-momentum tensor of a perfect fluid whose energy density and pressure are given by $\rho_m = (G - T)$ and $p_m = 0$ respectively. On the other hand, the mimetic field plays the role of velocity gradient for the fluid four-velocity: $v_{\mu} \equiv \partial_{\mu}\phi$. The equation of motion for the mimetic field can be obtained by varying the action with respect to the field itself, which gives \cite{mimetic1}:
\begin{eqnarray}
\nabla_{\mu}[(G - T)\partial^{\mu}\phi] = 0 \, ,
\label{mimeticeo}
\end{eqnarray}
where the covariant derivative is calculated with respect to the physical metric $g_{\mu \nu}$. \\

It is easy to show that, on a flat FLRW background, the energy density of the mimetic field scales with scale factor $a$ as $a^{-3}$, thus mimicking the contribution of pressureless dust \cite{mimetic1}. Hence, dark matter appears in mimetic gravity as a purely geometrical effect. Let us note that mimetic gravity differs from General Relativity by the appearance of an extra scalar degree of freedom.\footnote{However, mimetic gravity does not possess a proper scalar degree of freedom. Instead, the would-be scalar d.o.f. in mimetic gravity is constrained by Eq.(\ref{constraint}). It is easy to see that such constraint kills the wave-like parts of the would-be scalar d.o.f., implying that in the minimal mimetic gravity model the sound speed satisfies $c_s = 0$, and hence there are no propagating scalar degrees of freedom. Whether or not this leads to ghost instabilities is an issue which is far from being settled (see e.g. ~\cite{chaichian,kluson,firouzjahi,quintin,healthyimperfect,instabilities}), with different works reaching different conclusions concerning the stability of mimetic gravity and some of its basic extensions. In any case it appears that, if mimetic gravity does indeed suffer from serious ghost and/or gradient instability problems, these can nonetheless be cured by constructing a new type of curvature-invariant scalar function~\cite{quintin}, by non-minimally coupling the mimetic field to curvature~\cite{healthyimperfect} (note that we will explore the phenomenology of such a coupling in the context of rotation curves in this work), or by non-minimally higher derivatives of the mimetic field to curvature~\cite{instabilities}.} This can be traced to the fact that mimetic gravity is related to General Relativity by a singular disformal transformation~\cite{bek}: that is, the disformal transformation relating the physical to the auxiliary metric is non-invertible \cite{deruellerua,matarrese}. It is well known that this results in a theory possessing additional degrees of freedom \cite{yuan,domenech,deffayet,domenech1,langlois}.\footnote{See also \cite{ghalee1,carvalho,ghalee2} for recent work on the role of disformal transformations in theories of modified gravity.} \\

Finally, let us remark that the constraint on the mimetic field, Eq.(\ref{constraint}), can be enforced at the level of the action by means of a Lagrange multiplier term. That is, one can write the mimetic gravity action as \cite{mimetic2}:
\begin{eqnarray}
I = \int d^4x\sqrt{-g} \ \frac{1}{2\kappa^2} \left [ R + \lambda ( \partial^{\alpha}\phi\partial_{\alpha}\phi - 1 ) \right ] \, .
\label{lagrange}
\end{eqnarray}
In fact, it is easy to check that varying the action Eq.(\ref{lagrange}) with respect to the Lagrange multiplier $\lambda$ results precisely in the mimetic constraint Eq.(\ref{constraint}). The study of Lagrange multiplier-constrained theories, which represent the prototype for mimetic theories, had been initiated earlier in \cite{vikman,gao,capozziello}.\footnote{See also \cite{paliathanasis} for recent work on the role of Lagrange multipliers in cosmology.} Various aspects of mimetic gravity (including ghost and stability issues, null-energy condition violation, connections to other theories such as Ho\v{r}ava-Lifshitz gravity and the scalar Einstein-aether theory, and various solutions such as static spherically symmetric, cylindrical, stellar ones) and extensions thereof have been discussed at length in the quickly growing literature related to the subject \cite{golovnev,sheykin,momeni1,harko,
malaeb,battefeld,momenia,odintsov1,chamseddine,saadi,ramazanov,mirzagholi,leon,haghani,matsumoto,momeni2,
myrzakulov1,jacobson,speranza,astashenok1,myrzakulov2,myrzakulov3,bufalo,silva,momeni3,momeni4,khalifeh,ramazanov1,
nesseris,shiravand,raza,guendelman1,kan,odintsov2,guendelman2,rabochaya,oikonomou,nmyrzakulov,asenjo,myrzakulov4,
odintsov3,guendelman3,oikonomou1,sebastiani1,guendelman4,astashenok2,koshelev,ali,arroja,hammer,odintsov4,cognola,
myrzakulov5,chamseddine1,ramazanov2,nojiri,babichev,odintsov5,saridakis,weiner,sebastiani2,erices,ufd,saitou,shahidi,sebastianinp,
ijjas,oikonomou2,kopp,sebastiani3,sebastiani4,chamseddine2,vacaru,oscillations,kimura,viable,
denisa,aspects,guendelman5,continuous,focusing,matsumotorotations,chagoyatasinato,bhatia,qg1,qg2,elizalde,abebe,
singularities,nonsingular,guendelman6,bhatiaetal,tmsk,singh,axiomatization,
kleidisoikonomouapess,braneworld,wasay,bodendorfer,liu,causticfreebabichev,guendelman7,
cai,kamenshchik,mgreviewinanutshell,brevik,sahoo,baffou,singh2,cai2}, to which we refer the reader for further details. \\

Despite the huge efforts in the study of mimetic gravity, one important problem remains, at least to some extent, open. Namely, the question of whether it is possible to explain the flat galactic rotation curves in mimetic gravity. Two approaches to the problem have been considered or at least briefly discussed in the literature so far. Already in \cite{mimetic1} (see also \cite{ramazanov,mirzagholi}) it was recognized that, since mimetic dark matter behaves as dust, at least on cosmological scales, it is expected to be affected by gravitational instability: this consists of the usual gravitational collapse phenomenon leading to structure formation under the influence of gravity. Thus in principle, at sufficiently late times, mimetic dark matter could collapse and form virialized structures such as dark matter halos. However, exploring this possibility is highly non-trivial, given that at late times and on the relevant scales, mimetic dark matter is in the highly non-linear regime, much as particle dark matter in the analogous situation. Ultimately, a full understanding of the eventual nonlinear collapse in mimetic gravity would require powerful N-body simulations to be performed, leaving this path currently unexplored. \\

An alternative solution was proposed in \cite{myrzakulov4}, where it was realized that by choosing an appropriate potential for the mimetic field, it is possible to modify the effective Newtonian gravitational potential felt by a test particle, in particular by implementing a linear and quadratic correction to such potential. Several implementations of potentials were explored and it was shown that a satisfactory fit to rotation curves of spiral galaxies can in principle be obtained, at the cost of a quite contrived form for the potential. A related solution, also implemented through a potential for the mimetic field, was studied in~\cite{matsumotorotations}. Thus, the question of whether a simple and transparent explanation for flat rotation curves in mimetic gravity exists remains open. \\

In this \textit{Letter}, we shall explore yet an alternative solution. By considering a non-minimal coupling between the mimetic field and matter, we will show how, in the weak field limit, such coupling leads to the acceleration law which resembles a MOND-like one and thus can possibly satisfactorily explain the flatness of rotation curves.~\footnote{As we will show in the following section, a non-minimal coupling of matter to curvature leads to non-geodesic motion of test particles, as well as tiny violations of the equivalence principle. This is in principle expected to be tightly constrained by solar system tests. Surprisingly, it turns out that constraints on non-minimal couplings leading to non-geodesic motion of the form we will consider are subject to extremely loose constraints from non-cosmological observations (see e.g.~\cite{ep1,ep2,ep3,ep4,ep5,ep6,ep7,ep8,ep9,ep10}, and especially~\cite{ep11,ep12,ep13,ep14}). In many cases, however, the loose constraints are a result of the only partial applicability of the applied methodology, which thus calls for a revisiting. Therefore, in the upcoming work \cite{mine}, together with a thorough examination of rotation curves data, we plan to more carefully revisit constraints on this type of non-minimal coupling from solar system tests.}

\section{Non-minimal coupling of the mimetic field}
\label{nm}

Let us now introduce a non-minimal coupling between the mimetic field and the matter hydrodynamic flux, of the form:
\begin{eqnarray}
\gamma J^{\mu}\partial_{\mu}\phi \, ,
\label{minimal}
\end{eqnarray}
where the matter hydrodynamic flux current is given by:
\begin{eqnarray}
J^{\mu} = \rho u^{\mu} \, ,
\label{current}
\end{eqnarray}
$\rho$ being the energy density of the matter content of the Universe, which we model as a perfect fluid. Similar or analogous couplings have been considered recently in the literature, see e.g. \cite{dominance,francaviglia,extra,tskoivisto}. Therefore the action for mimetic gravity is modified to the following:
\begin{eqnarray}
I = \int d^4x\sqrt{-g} \ \frac{1}{2\kappa^2} \left [ R + \lambda ( \partial^{\alpha}\phi\partial_{\alpha}\phi - 1 ) + {\cal L}_m \right ] \, , \nonumber \\
\end{eqnarray}
where the non-minimal coupling between matter and mimetic sector is included in the matter Lagrangian ${\cal L}_m$, which reads:
\begin{eqnarray}
{\cal L}_m = -\rho + \gamma J^{\mu}\partial_{\mu}\phi \, .
\label{lagrangian}
\end{eqnarray}
With the definition as per Eq.(\ref{lagrangian}), the energy density of matter (understood to be the $00$ component of the energy-momentum tensor, proportional to the functional derivative of the matter Lagrangian with respect to the physical metric) can be unambiguously identified with $\rho$, thus allowing for a proper definition of the non-minimal coupling in Eq.(\ref{minimal}), through Eq.(\ref{current}). \\

Mathematical as well as more phenomenological considerations restrict the range of allowed values of the strength of the non-minimal coupling $\gamma$, see the very relevant analysis conducted in~\cite{ep10,dutta}. In particular stability and causality considerations restrict $\gamma \geq 0$. An upper bound on $\gamma$ can be set through solar system tests, which constrain non-geodesic motion and violations of the equivalence principle. As mentioned earlier (see footnote 5), this requires revisiting previous bounds on such theories which were found to be quite loose: carrying out such an analysis here would be beyond the scope of this work and hence we choose to include it in a follow-up paper where we include an analysis of spiral galactic rotation curves data~\cite{mine}. \\

The equations of motion of the theory can be derived by varying the action with respect to the metric, the mimetic field, and the Lagrange multiplier. Variation with respect to the metric leads to the gravitational field equations:
\begin{eqnarray}
G_{\mu \nu} = \kappa ^2T_{\mu \nu} - \frac{\gamma}{2}T_{\mu \nu}u^{\sigma}\partial_{\sigma}\phi - \lambda \partial_{\mu}\phi\partial_{\nu}\phi \, .
\label{einstein}
\end{eqnarray}
The equations of motion for the mimetic field can be derived by varying the action with respect to the latter, and read:
\begin{eqnarray}
\nabla_{\mu}(\lambda \nabla^{\mu}\phi) = -\frac{\gamma}{2}\nabla_{\mu}(\rho u^{\mu}) \, ,
\label{gravitational}
\end{eqnarray}
which can be rewritten as a conservation equation, i.e. $\nabla_{\mu}{\cal J}^{\mu} = 0$, where the covariantly conserved current is:
\begin{eqnarray}
{\cal J}^{\mu} = \lambda \partial^{\mu}\phi + \frac{\gamma}{2}\rho u^{\mu} \, .
\label{field}
\end{eqnarray}
Finally, variation of the action with respect to the Lagrange multiplier enforces the mimetic constraint, Eq.(\ref{mimeticc}):
\begin{eqnarray}
\partial ^{\mu}\phi\partial_{\mu}\phi = 1 \, .
\label{constraint}
\end{eqnarray}
Thus, the equations of motion of the theory are given by Eqs.(\ref{einstein},\ref{gravitational},\ref{field},\ref{constraint}). \\

Because of the non-minimal coupling between the matter hydrodynamic flux and the mimetic field, we expect the matter energy-momentum tensor to no longer be covariantly conserved, i.e. $\nabla_{\mu}T^{\mu \nu} \neq 0$. To show this explicitly, let us model the matter content of the Universe as that of a perfect fluid whose energy density and pressure are denoted by $\rho$ and $p$ respectively. Then, the matter energy-momentum tensor is given by:
\begin{eqnarray}
T_{\mu \nu} = (\rho + p)u_{\mu}u_{\nu} + pg_{\mu \nu} \, ,
\label{energymomentumtensor}
\end{eqnarray}
where $u^{\mu}$ denotes the four-velocity of the fluid and is the quantity appearing in Eq.(\ref{current}). Then, a length but straightforward calculation shows that:
\begin{eqnarray}
\nabla^{\mu}T_{\mu \nu} = \frac{\gamma \left [ T_{\mu \nu}\nabla ^{\mu}(u^{\sigma}\partial_{\sigma}\phi) - \rho \nabla_{\sigma}u^{\sigma}\partial_{\nu}\phi \right ]}{2\kappa ^2 - \gamma u^{\sigma}\partial_{\sigma}\phi} \, .
\label{nonconservation}
\end{eqnarray}
Of course $\gamma \neq 2\kappa^2/(u^{\mu}\partial_{\mu}\phi)$ is required in order for the denominator of Eq.~\ref{nonconservation} not to vanish and hence lead to divergences. As expected, in the presence of the non-minimal coupling between matter and mimetic field, the energy-momentum tensor of matter is no longer covariantly conserved. This is not unexpected physically, given that it corresponds to an exchange of energy and momentum between two sectors: the ordinary matter sector, and the mimetic sector (responsible for the mimetic dark matter). \\

To test the motion of particles in the presence of the non-minimal coupling, we need to determine how the usual geodesic equation, $u^{\mu}\nabla_{\mu}u^{\nu} = 0$, is modified in the presence of such coupling. It is useful to define the projection operator $h_{\mu \nu} = g_{\mu \nu} + u_{\mu}u_{\nu}$. Then, it is immediate to verify that $h_{\mu \nu}$ satisfies $h_{\mu \nu}u^{\mu} = 0$. The divergence of Eq.(\ref{energymomentumtensor}), after raising both indices of the energy-momentum tensor, is easy to express by making use of the projection operator:
\begin{eqnarray}
\nabla_{\mu}T^{\mu \nu} &=& h^{\mu \nu}\partial_{\mu}p + u^{\nu}u_{\mu}\partial^{\mu}\rho \nonumber \\
&+& (\rho + p)(u^{\nu}\nabla_{\mu}u^{\mu} + u^{\mu}\nabla_{\mu}u^{\nu})
\label{divergence}
\end{eqnarray}
We can now contract Eq.(\ref{divergence}) with $h^{\sigma}_{\nu}$, yielding:
\begin{eqnarray}
h^{\sigma}_{\nu}\nabla_{\mu}T^{\mu \nu} = (\rho + p)u^{\mu}\nabla_{\mu}u^{\sigma} + h^{\nu \sigma}\partial_{\nu}p \, .
\label{contract}
\end{eqnarray}
To proceed, recall that one can express $u^{\mu}\nabla_{\mu}u^{\sigma}$, i.e. the parallel transport of the matter content fluid four-velocity, by making use of the Christoffel symbols as:
\begin{eqnarray}
u^{\mu}\nabla_{\mu}u^{\sigma} = \frac{d^2x^{\sigma}}{ds^2} + \Gamma^{\sigma}_{\mu\nu}u^{\mu}u^{\nu} \, ,
\label{parallel}
\end{eqnarray}
where as usual $ds^2$ denotes the (squared) proper line element. \\

By combining Eqs.(\ref{divergence},\ref{contract},\ref{parallel}), we obtain the generalized geodesic equation in the presence of the non-minimal coupling between matter and mimetic field, which takes the form:
\begin{eqnarray}
\frac{d^2x^{\sigma}}{ds^2} + \Gamma^{\sigma}_{\mu \nu}u^{\mu}u^{\nu} = f^{\sigma} \, ,
\label{geodesic}
\end{eqnarray}
with $f^{\sigma}$ given by:
\begin{eqnarray}
f^{\mu} = \frac{h^{\mu\sigma}}{\rho + p} \left ( \frac{\gamma[\nabla_{\sigma}(p u^{\alpha}\partial_{\alpha}\phi) - \partial_{\sigma}\phi\nabla_{\alpha}(\rho u^{\alpha})] - 2\kappa^2\partial_{\sigma}p}{2\kappa^2 - \gamma u^{\alpha}\partial_{\alpha}\phi} \right ) \nonumber \\
\label{force} 
\end{eqnarray}
As we had anticipated, the presence of the non-minimal coupling, which leads to energy and momentum transfer between matter and mimetic field, also leads to non-geodesic motion (with geodesic motion corresponding to $f^{\mu} = 0$). In fact, the non-minimal coupling leads to the appearance of an extra non-geodesic force, $f^{\mu}$. Importantly, because of the appearance of the projection operator $h^{\mu\sigma}$ in Eq.(\ref{force}), the non-geodesic force is perpendicular to the matter four-velocity, i.e. $f^{\mu}u_{\mu} = 0$. In the upcoming subsection we shall explore the physical properties of such force and, correspondingly, its impact on the rotation curves of spiral galaxies.

\section{Newtonian limit}
\label{nl}

To make progress and understand the effects of the non-geodesic force, we consider the Newtonian limit of Eq.(\ref{force}). Let us define the total acceleration of a test particle as $a^{\mu} \equiv d^2x^{\mu}/ds^2$. Thus the spatial part of $a^{\mu}$ is given by $\vec{a} = (a^1,a^2,a^3)$, where $a^i = d^2x^i/ds^2$. Similarly, let us define $a_N^{\mu}$, the four-acceleration encoding ordinary gravitational effects, as $a_N^{\mu} \equiv -\Gamma^{\mu}_{\alpha\beta}u^{\alpha}u^{\beta}$. The spatial part of $a_N^{\mu}$ is given by $\vec{a}_N = (a_N^1,a_N^2,a_N^3)$, where $a_N^i = -\Gamma^{i}_{\alpha\beta}u^{\alpha}u^{\beta}$. For a point mass distribution, $\vec{a}_N = GM\hat{r}/r^3$. Finally, let us denote by $\vec{f}$ the spatial part of the non-geodesic force $f^{\mu}$. \\

We can express the Newtonian limit of the generalized geodesic equation Eq.(\ref{geodesic}) as:
\begin{eqnarray}
\vec{a} = \vec{a}_N + \vec{f} \, .
\label{newtonian}
\end{eqnarray}
In other words, the total acceleration is given by two contributions: the usual Newtonian acceleration arising in the weak field limit of GR, and a non-geodesic contribution. By squaring Eq.(\ref{newtonian}) we obtain:
\begin{eqnarray}
a^2 = a_N^2 + f^2 + 2\vec{a}_N \cdot \vec{f} \implies \vec{f} \cdot \vec{a}_N = \frac{1}{2}(a^2 - a_N^2 - f^2) \, . \nonumber \\
\label{dot}
\end{eqnarray}
It is straightfoward to verify that from Eq.(\ref{dot}) one can express $\vec{a}_N$ as:
\begin{eqnarray}
\vec{a}_N = \frac{1}{2}(a^2 - a_N^2 - f^2)\frac{\vec{a}}{\vec{f} \cdot \vec{a}} \, .
\label{an}
\end{eqnarray}
Notice also that Eq.(\ref{an}) requires the non-geodesic force and the total acceleration to not be orthogonal, i.e. $\vec{a} \cdot \vec{f} \neq 0$. For simplicity, let us consider the case where the two are parallel, i.e. $\vec{f} \cdot \vec{a} = fa$. Considering the two to not be parallel does not affect our conclusions sensibly. Therefore, the Newtonian acceleration $\vec{a}_N$ in Eq.(\ref{an}) can be expressed as:
\begin{eqnarray}
\vec{a}_N = \frac{1}{2}(a^2 - a_N^2 - f^2)\frac{\vec{a}}{fa} \, ,
\label{annew}
\end{eqnarray}
that is, the Newtonian acceleration is parallel to the total acceleration. \\

To explore the impact of the non-geodesic force on galactic rotation curves, it is useful to consider the limit where the usual gravitational acceleration is negligible, i.e. $a_N \ll a$. Recall that, in the region of approximate flatness of galactic rotation curves, the contribution of baryonic matter is negligible with respect to the contribution supposedly coming from the dark matter halo. In this limit, Eq.(\ref{annew}) is modified to the following:
\begin{eqnarray}
\vec{a}_N \simeq \frac{1}{2} \left ( \frac{a}{f} - \frac{f}{a} \right ) \vec{a} \, .
\label{weak}
\end{eqnarray}
The above Eq.(\ref{weak}) assumes a more familiar connotation if we define the following:
\begin{eqnarray}
a_0 \equiv \frac{2fa^2}{a^2 - f^2} \, .
\label{a0}
\end{eqnarray}
Then, one finds that:
\begin{eqnarray}
\vec{a}_N \simeq \frac{a}{a_0}\vec{a} \, ,
\label{m}
\end{eqnarray}
which trivially follows from combining Eqs.(\ref{weak},\ref{a0}). \\

The above Eq.(\ref{m}) bears striking resemblance with the acceleration equation in MOND, a proposed modification to Newtonian mechanics where Newton's law is modified at very small accelerations, corresponding to those experienced by point particles at the very outer edges of galaxies \cite{milgrom} (see also \cite{galaxysystems,appb}). In fact, in order to match the inferred rotation curves, the parameter $a_0$ is estimated to be of the order of ${\cal O}(10^{-10}) m/s^2$. It is worth noticing that the MOND regime has also been recovered in the weak-field limit of a particular extended metric theory of gravity (see e.g. \cite{etg,delacapo} for reviews concerning the appearance of dark matter in extended theories of gravity), wherein the acceleration scale $a_0$ is treated as a fundamental parameter which breaks scale invariance in gravitational interactions \cite{bernal}, making gravitational interactions scale dependent (see also \cite{troisi,cardone,salzano,sobouti}). \\

To see how the approximate flatness of galactic rotation curves can be inferred from Eq.(\ref{m}), consider a point particle at a distance $r$ from the center of a galaxy of mass $M$, sufficiently far from the region where baryonic contributions to the gravitational potential are dominant. Then the Newtonian contribution to the acceleration equation is given by:
\begin{eqnarray}
a_N = \frac{GM}{r^2} \, .
\end{eqnarray}
On the other hand, Eq.(\ref{m}) can be expressed as the following:
\begin{eqnarray}
a \simeq \sqrt{a_0a_N} = \sqrt{\frac{a_0GM}{r^2}} \, .
\label{cen}
\end{eqnarray}
Furthermore, the centripetal acceleration of a point particle in circular motion around a body of mass $M$ is given by $a = v_{\text{rot}}^2/r$. Thus, from Eq.(\ref{cen}) we infer that the rotational velocity of a point particle in the outskirts of a galaxy is approximately constant:
\begin{eqnarray}
v_{\text{rot}} \simeq \sqrt{a_0GM} \, ,
\end{eqnarray}
which is to be contrasted with the usual Newtonian result where $v_{\text{rot}} \propto 1/\sqrt{r}$. As in MOND, if we further assume that the luminosity of a galaxy is proportional to its mass, one immediately infers that $L \propto v_{\text{rot}}^4$, an empirical relation known as the Tully-Fisher relation. \\

A comment is in order here: for the full MOND regime to be recovered, $a_0$ in Eq.(\ref{a0}) should technically speaking be a constant. To see whether this is the case, or within what conditions $a_0$ can be considered constant, let us go ahead and find an explicit expression for $a_0$, starting from Eqs.(\ref{a0},\ref{cen}). Specifically, we obtain:
\begin{eqnarray}
a_0 = \frac{2fa^2}{a^2-f^2} = \frac{2f}{1-\frac{f^2}{a^2}} = \frac{2f}{1-\frac{f^2r^2}{a_0GM}} \, ,
\end{eqnarray}
which implies that:
\begin{eqnarray}
a_0 \left ( 1-\frac{f^2r^2}{a_0GM} \right ) = 2f \, ,
\end{eqnarray}
finally leading to:
\begin{eqnarray}
a_0(r) = 2f + \frac{f^2r^2}{GM} \, ,
\label{a0r}
\end{eqnarray}
which is our final expression for the acceleration scale in non-minimally coupled mimetic gravity as a function of the extra force. \\

The above Eq.(\ref{a0r}) shows that the acceleration scale $a_0$ is not precisely a constant, but carries an $r$-dependence, the exact details of which depend on the radial profile of the extra force. Therefore, the weak field limit of non-minimally coupled mimetic gravity does not exactly recover the full MOND regime, but rather a MOND-like one. Before discussing the very interesting consequences of this observation, let us comment on a caveat to the previous conclusion: namely, that under specific conditions, $a_0$ can be considered constant. In particular, if the extra force carries a radial dependence of the form $f(r) \propto 1/r$, then it is easy to see that in the limit where $r \rightarrow \infty$, $a_0$ approaches a constant.~\footnote{It is not obvious from Eq.(\ref{force}) whether in the Newtonian limit the extra force actually decays as $1/r$. If it did, as explained above, then non-minimally coupled mimetic gravity would exactly recover the MOND regime. Even if it does not, nonetheless, non-minimally coupled mimetic gravity still recovers a MOND-like regime which can have extremely interesting phenomenological applications, especially in view of some of the weaknesses of MOND, as explained in the following paragraph. To check the exact functional form of the extra force in the Newtonian limit, $f(r)$, is a complex task which would bring us beyond the scope of this \textit{Letter}, since it requires a precise modelling of the density and pressure profiles of galaxies on a vast range of scales, an issue which has yet to reach consensus even on the observational side. We will return to this issue in detail in~\cite{mine}.} \\

Let us now comment on the consequences of $a_0$ carrying a radial dependence. In principle this possible weakness of the model can be turned into a surprising strength. Specifically, it can be used to overcome a weakness of MOND, namely the fact that it can only account for observations on galactic scales. The reason for this resides in the fact that MOND only carries a single constant acceleration scale, which is tuned to fit rotation curves of galaxies (particularly medium-sized spiral galaxies). However, the acceleration scale which a MOND-like theory would require in order to fit increasingly large or small astrophysical objects is correspondingly increasingly larger or smaller, and thus cannot be accommodated by a single fixed scale. \\

In non-minimally coupled mimetic gravity, depending on the properties of the extra force and in particular its radial dependence (e.g. whether $f(r)$ is a growing function of $r$, or decreases no faster than $1/r$), the acceleration scale $a_0$ can take the form of a growing function of $r$. In other words, the acceleration scale can in principle be tuned to vary with $r$ in such a way to reproduce the correct phenomenology both on galactic and cluster scales, which cannot be achieved in the simplest realization of MOND. Therefore, non-minimally coupled mimetic gravity can be viewed as a covariant realization of a MOND-like theory which can address the missing mass problem on multiple scales, not only galactic ones. This is an observation which of course deserves further investigation, including a detailed modelling of the extra force which of course depends on the density and pressure profiles of the astrophysical objects in question: because such a detailed study would bring us beyond the scope of this \textit{Letter}, we choose to include it in our upcoming work \cite{mine}. \\

Finally, let us make an important remark concerning Eq.(\ref{a0r}) on spiral galaxy scales, \textit{i.e.} those of interest for this \textit{Letter}. Namely, the spiral galaxy zoo is surprisingly uniform in terms of physical properties such as density profiles, masses, sizes, and asymptotic rotational velocities (of the order of $200 \ {\rm km/s}$). Therefore the physical properties determining the properties of the extra force Eq.(\ref{force}) as well as the acceleration scale Eq.(\ref{a0r}) are quite uniform across the distribution of medium-sized spiral galaxies, implying indeed that the acceleration scale $a_0$ on the scale of medium-sized spiral galaxies can be considered approximately constant. As we discussed previously, its evolution with scale can instead be exploited to try and solve the missing mass problem on multiple scales, e.g. galactic and cluster ones. \\

An immediate corollary of our findings, with caveats due to the approximations discussed above, is that in mimetic gravity the successes of MOND in fitting spiral galaxy rotation curves are expected to be recovered \cite{radialacceleration} (see also \cite{successes1,successes2,successes3,successes4,successes5,successes6}) through the non-minimal coupling. Of course, to ultimately confirm that the successes of MOND can indeed be reproduced within the framework of non-minimally coupled mimetic gravity, it is mandatory to test the theory on rotation curves data. A discussion on fit to rotation curves would bring us beyond the initial scope of this \textit{Letter}, so we have decided to reserve this work to a companion paper which will appear later \cite{mine}. \\

\section{Conclusions}
\label{conclusion}

In this \textit{Letter}, we have shown how a non-minimal coupling between the mimetic field and the matter hydrodynamic flux in mimetic gravity generates a non-geodesic force term which, in the Newtonian limit, leads to a MOND-like acceleration law. In this way, we expect it to be possible to reproduce the successes of MOND within the framework of non-minimally coupled mimetic gravity. This further suggests that non-minimally coupled mimetic gravity can be viewed as a covariant realization of a MOND-like theory. \\

There is ample avenue for further work on the topic. Firstly, it is interesting to fully investigate whether the model in question can overcome the challenges of MOND on cluster scales. We have suggested a mechanism which has the potential to address this issue, which exploits the scale-dependence of the acceleration scale recovered in non-minimally coupled mimetic gravity. Of course this explanation requires more robust investigation in order to achieve validation. Next, and most importantly, it is mandatory to test the theory on data: in other words, to adequately explore the region of parameter space within which the rotation curves of galaxies can be satisfactorily fitted, with rigorous statistical methods. Finally, it is also important to carefully revisit bounds on non-minimally coupled mimetic gravity from solar system tests. These and other issues will be addressed in a companion paper \cite{mine}. \\

\begin{acknowledgments}
I thank the Niels Bohr Institute, where most of this work was completed and written, for hospitality. Part of this work was support by the Vetenskapsr\r{a}det (Swedish Research Council). I am grateful to Salvatore Capozziello, Amel Durakovi\'{c}, Zahra Haghani, Purnendu Karmakar, Sabino Matarrese, Sergei Odintsov, Massimiliano Rinaldi, Emmanuel Saridakis, Lorenzo Sebastiani, Shahab Shahidi, and Sergio Zerbini for very useful discussions which led to this work, and during its preparation.
\end{acknowledgments}

%

\end{document}